\documentclass[letter]{aa} 

\usepackage{graphicx}
\usepackage{txfonts}
\begin{document}

   \title{Dust Properties of Double-Tailed Active Asteroid (6478) Gault}

   \subtitle{}
                          
   \author{
F. Moreno\inst{1}\and
E. Jehin \inst{2}\and
J. Licandro\inst{3,4}\and
M. Ferrais \inst{2}\and
Y. Moulane \inst{2,6,8}\and
F.J. Pozuelos \inst{2,5}\and
J. Manfroid \inst{2}\and
M. Devog\`ele \inst{7}\and
Z. Benkhaldoun \inst{6}\and
N. Moskovitz \inst{7}\and
M. Popescu\inst{3,4}\and
M. Serra-Ricart\inst{3,4}\and
A. Cabrera-Lavers\inst{9,3,4}\and
M. Monelli\inst{3,4}
}

   \institute{Instituto de Astrof\'\i sica de Andaluc\'\i a, CSIC, Glorieta
    de la Astronom\'\i a S/N, 18008 Granada, Spain\\
              \email{fernando@iaa.es}
              \and
    Space sciences, Technologies \& Astrophysics Research (STAR) Institute, Universit\'e de Li\`ege, 4000 Li\`ege, Belgium \and            
 Instituto de Astrof\'\i sica de Canarias, V\'\i a L\'actea s/n, 38205, 
 La Laguna, Spain \and
 Departamento de Astrof\'\i sica, Universidad de 
 La Laguna, 38206, La Laguna, Tenerife, Spain
\and
  EXOTIC Lab, UR Astrobiology, AGO Department, University of Li\`ege, 4000 Li\`ege, Belgium \and
  Oukaimeden Observatory, High Energy Physics and Astrophysics Laboratory, Cadi Ayyad University, Marrakech, Morocco\and
  Lowell Observatory, 1400 West Mars Hill Road, Flagstaff, AZ 86001 (U.S.A.)\and
  ESO (European Southern Observatory) - Alonso de Cordova 3107, Vitacura, Santiago Chile\and
  GRANTECAN, Cuesta de San Jos\'e s/n, E-38712 Bre\~na Baja, La Palma,
  Spain }


 
  \abstract
{ Asteroid (6478) Gault was discovered to exhibit a comet-like tail in
  observations from December 2018, becoming a new member of the
  so-called active asteroid population in the main asteroid belt.} 
   { The aims are to investigate the grain properties
     of the dust ejected from asteroid (6478) Gault and to give insight
     into the activity mechanism(s).}
   {We use a Monte Carlo dust tail brightness code to retrieve the
     dates of dust ejection, the physical properties of the grains,
     and the total dust mass losses during each event. The code takes
     into account the brightness contribution of the asteroid
     itself. The model is applied to a large data set of images
     spanning the period from January 11, 2019 to March 13, 2019. In
     addition, both short- and long-term photometric measurements
     of the asteroid have been carried out.}
   {It is shown that, to date, asteroid (6478) Gault has experienced two
     episodes of impulsive dust ejection, that took place around 2018
     November 5 and 2019 January 2, releasing at least
     1.4$\times$10$^7$ kg and 1.6 
     $\times$10$^6$ kg of dust, respectively, at escape speeds. The size
     distribution, consisting of particles in the 1 $\mu$m to 1 cm
     radius range, follows a broken power-law with bending
     points near 15 $\mu$m and 870 $\mu$m. On the other hand, the
     photometric series indicate a nearly constant magnitude over
     several 5--7.3 h periods, a possible effect of the masking of a
     rotational lightcurve by the dust.} 
   {The dust particles forming Gault's tails were released from the
     asteroid at escape speeds, but the specific ejection mechanism is
     unclear until photometry of the dust-free asteroid are conducted, in
     order to assess whether this was related to rotational disruption or
   to other possible causes.}


   \maketitle
%

\section{Introduction}

 Main belt asteroid (6478) Gault (hereinafter Gault for short), a
 member of the 25 Phocaea asteroid family (Small Bodies Data Ferret,
 NASA Planetary Data System, see https://sbnapps.psi.edu/olaf/) was
 reported to have a comet-like tail at PA=290$^\circ$
on December 8, 2018, from ATLAS 
 images, and early Finson-Probstein (Finson \& Probstein, \citeyear{Finson}) 
analysis of those 
 images revealed ejection of material in early November 2018
 (Smith et al. \citeyear{Smith19}). 
Previous images from ATLAS and PanSTARRS back to 2010 do not
 show any sign 
 of activity.  Follow-up imaging of the object revealed the presence of another
 growing tail-like structure at about PA=305$^\circ$ (Jehin et al. 
 \citeyear{Jehin19}), probably associated to a second ejection event.

The orbital elements of Gault are $a$=2.305 au, $e$=0.194, and
$i$=22.81$^\circ$, so 
 it is an inner main belt object with a Tisserand parameter with respect to
 Jupiter of $T$=3.46. This 
 classifies Gault as a new member of the active asteroid population
 (e.g., Jewitt et al. \citeyear{Jewitt15}). As with many other active
 asteroids that 
 show long-term dynamical stability (e.g. Haghighipour
 \citeyear{Haghighipour09}, Jewitt et al.  
 \citeyear{Jewitt09}), its 
 orbit is stable on timescales of 100 Myr or longer, as we have
 verified by backward dynamical evolution using the Mercury integrator 
 (Chambers \citeyear{Chambers99}) starting from the current orbital
 elements of Gault. This implies that Gault is a native member of the main
 belt. Among the possible causes for 
 the activity in these objects, impact, rotational breakup, and ice
 sublimation (or even a combination of these) are usually invoked
 (see Jewitt et al. \citeyear{Jewitt15} for a complete description of
 the possible mechanisms involved).  In principle,
 ice-sublimation-related activity on Gault seems 
 unlikely owing to its character of inner belt asteroid. Likewise,
 the occurrence of two succesive impacts seems rather improbable.

 The appearance of Gault is 
 becoming similar to asteroid 311P (Jewitt et al. \citeyear{Jewitt13}, Moreno et
 al. \citeyear{Moreno14}, Hainaut et al. 
 \citeyear{Hainaut14}), currently described as  a close binary in which
 one of the components  rotates near the centripetal limit (Jewitt et
 al. \citeyear{Jewitt18}).

   \begin{figure*}
   \centering
   \includegraphics[angle=0,width=15cm]{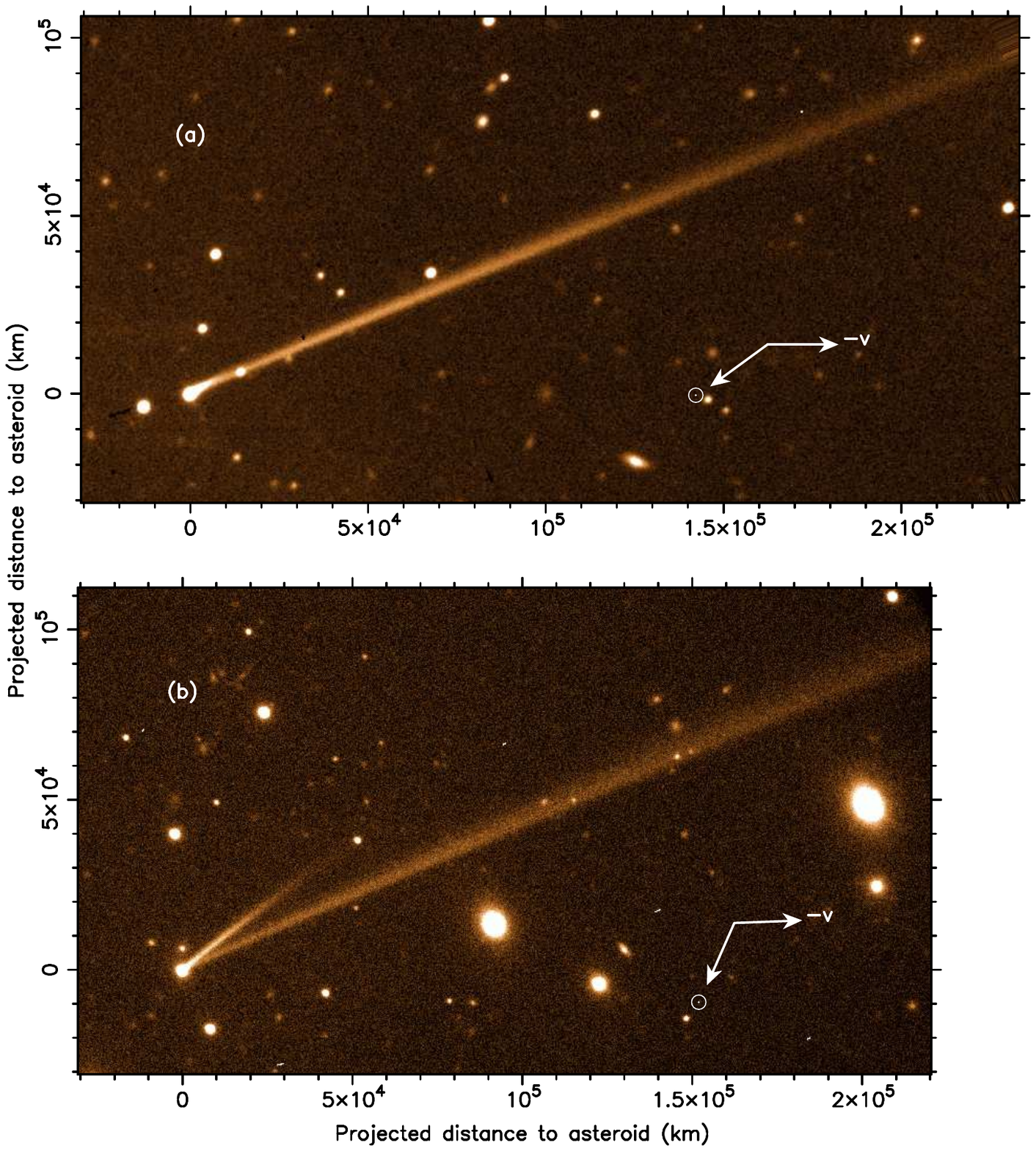}
   \caption{Sample images of asteroid (6478) Gault. Panel (a) shows
     the image obtained on January 14, 2019, with OSIRIS at the 10.4m
     Gran Telescopio Canarias, through a  Sloan $r$ filter.  Panel (b)
     shows an image obtained on February 15, 2019, with SOAR 4.1m telescope
     through a $V+R_c$ broadband filter. North is up, East to the
     left.  The directions to the 
     Sun and the negative of the heliocentric velocity vector (v) are shown.}
              \label{Fig1}%
    \end{figure*}
 
 In this paper, we describe observations and
 models of Gault to impose 
 constraints on the physical properties of the ejected dust, and to
 shed light on the activity timeline and causes for the ejection events.     

\section{Observations and data reduction}

   Images of Gault have been obtained at
   various telescopes around the world in the time frame from January
   11 to February 
   15, 2019. Table~\ref{log} shows the log of the observations. Except for the
   observations from SOAR 4.1m telescope, in which a broadband $V+R_c$ 
   filter was 
   used, either Sloan $r$ or $R_c$ (Cousins) filters were used. The
   images from GTC 
   10.4m telescope were calibrated using standard stars, while in all
   the other cases, field stars and the USNO-B1.0 catalogue was used
   in the reduction procedure. Median stacking was performed for the
   TRAPPIST (Jehin et al. \citeyear{Jehin11}) images to improve the signal-to-noise ratio. The seeing
   disk varied between 0.8 and 1.5 arcsec FWHM.

   \begin{table*}
      \caption{Journal of the observations: imaging.}
         \label{log}
\centering                          
\begin{tabular}{l c c c c c}        
\hline\hline                 
Telescope/Instrument & Date (UT) & $r$ & $\Delta$ & $\alpha$ & Scale \\   
                     &      & (au) & (au) & (deg) & arcsec/pixel \\
\hline                        
TCS/MuSCAT2 & 2019-01-11T02:43 & 2.464 & 1.822 & 20.3 & 0.434 \\
GTC/OSIRIS  & 2019-01-13T06:56 & 2.460 & 1.794 & 19.8 & 0.254 \\
GTC/OSIRIS  & 2019-01-14T06:12 & 2.458 & 1.781 & 19.7 & 0.254 \\
TRAPPIST-S    & 2019-01-21T00:11 & 2.444 & 1.699 & 18.1 & 1.250 \\
TRAPPIST-S    & 2019-01-29T05:36 & 2.428 & 1.608 & 15.9 & 1.250 \\
TRAPPIST-S    & 2019-02-05T02:45 & 2.414 & 1.541 & 13.7 & 1.250 \\
TRAPPIST-S    & 2019-02-07T07:22 & 2.410 & 1.523 & 12.9 & 0.625 \\
SOAR        & 2019-02-15T06:37 & 2.394 & 1.463 & 10.1 & 0.290 \\
TRAPPIST-S    & 2019-03-05T05:34 & 2.356 & 1.388 &  6.9 & 0.625 \\
TRAPPIST-S    & 2019-03-13T04:48 & 2.338 & 1.384 &  8.8 & 0.625 \\
\hline                                   
\end{tabular}
\tablefoot{Telescope acronyms:
  TCS: 1.52m Carlos S\'anchez Telescope at Tenerife. GTC:
  10.4m Gran Telescopio Canarias. TRAPPIST: 0.6m Transiting Planets and
  Planetesimals Small Telescope. SOAR: 4.1m Southern Astrophysical
  Research Telescope. $\Delta$ and $r$ are the asteroid geocentric and
  heliocentric  distances, respectively, and $\alpha$ is the phase angle.} 
\end{table*}

The early images show a single  narrow tail near PA=290$^\circ$,
and a dense nuclear 
   condensation (Figure 1a). Since early February, a second tail started to
   show up near PA=305$^\circ$ (Figure 1b).

  In addition to the images, long photometric series were
  obtained with both 
TRAPPIST-North (TN) and -South (TS) on 2019 January 13, 14 and 15, in
order to determine the rotation period of the asteroid. In total we acquired 5
lightcurves of about 5 hours each, taken with the $R_c$ filter and an
exposure time of 120 s (Table 2). On January 14 and 15, we observed first with TN and then with TS, all with
the $R_c$ filter, which
allowed continuous observation during 7.3 hours, with some overlap between
the two telescopes. The calibration of the images was made with IRAF
scripts using corresponding flat fields, bias and dark frames. The
photometry was derived using the PhotometryPipeline (Mommert
\citeyear{Mommert17}). About 700 to 900  
stars in the PanSTARRS catalogue were used for the photometric
calibration to obtain the apparent $R_c$ magnitudes.

\begin{table}[htpb!]
    \centering
    \caption{Summary of Gault short-term photometric series ($R_C$ filter). }
\begin{tabular}{ c c c c c c}
   Date & $N_p$ & $r$ & $\Delta$    & $\alpha$ &  Site \\
   (UT)      &       & (au)   & (au) & (deg)     &        \\
  \hline
  \hline
  2019-01-13.3 & 119 & 2.46 & 1.79 & 19.9 &  TRAPPIST-S \\
  2019-01-14.2 & 124 & 2.46 & 1.78 & 19.7 &  TRAPPIST-N \\
  2019-01-14.3 & 134 & 2.46 & 1.78 & 19.7 &  TRAPPIST-S \\
  2019-01-15.2 & 124 & 2.45 & 1.77 & 19.5 &  TRAPPIST-N \\
  2019-01-15.3 & 131 & 2.45 & 1.77 & 19.4 &  TRAPPIST-S \\
  \hline
\end{tabular}
\end{table}

\section{The Dust Tail Brightness Code}

   In the simulations of the tails brightnesses, we used our Monte
   Carlo dust tail code that has been
   already described in several works in the past to characterise the dust
   environments of comets and main-belt comets (see
   e.g. Moreno et al. \citeyear{Moreno16}, Moreno et
   al. \citeyear{Moreno17}), so that 
   only a brief description is given. The dynamics of the particles
   (assumed spherical) are described by the $\beta$ parameter, defined
   as $\beta = \frac{C_{pr}Q_{pr}}{\rho d}$, where $d$ is the particle
   diameter, $C_{pr}$=1.19$\times$10$^{-4}$ g cm$^{-2}$ is the radiation
   pressure coefficient, and $Q_{pr}$ is the scattering efficiency for
   radiation pressure, which is $Q_{pr} \approx$1 for absorbing
   particles large compared to the observation wavelength (Burns
   \citeyear{Burns79}).    

  As Gault is an inner 
   belt asteroid of likely S-type, 
   the particle density is assumed at $\rho$=3400 kg
   m$^{-3}$, which is appropriate for S-type asteroid Itokawa regolith 
     (Tsuchiyama et al. \citeyear{Tsuchiyama11}). The code computes
   the position in the sky plane of a 
     large number of particles emitted isotropically from the asteroid, whose
     trajectories depend on $\beta$ and their terminal velocities
     (e.g., Fulle \citeyear{Fulle89}). We assume a broad differential
     size distribution of 
     particles between 1 $\mu$m and 1 cm in radius, following a
     power-law function.

   Prior to this activity period, Gault absolute magnitude has been
   reported as $H$=14.4 (JPL Small-Body Database). We assume a geometric
   albedo of $p_v$=0.15, appropriate for a S-type asteroid (e.g. Luu
     \& Jewitt \citeyear{LuuJewitt89}). Then, its
   diameter can be calculated as 
   $D$=4.5 km, using the equation by \citet{HarrisLagerros02}. Owing
   to its significant size, we added in the simulations the asteroid
   brightness contribution to the tail 
   brightness, assuming a constant cross section with time.  To
   produce
   realistic model images to be compared with
   the observations, we performed a convolution of the synthetic
   images obtained with a two-dimensional Gaussian function. The
   full-width at half-maximum of the Gaussian was set to the average 
   seeing value at the corresponding
   observation date. The asteroid and the dust particles brightness
   were corrected for the effect of the phase angle. For the asteroid we   
   applied a linear phase coefficient of $b$=0.033 mag deg$^{-1}$, which
   is computed from the relation by 
   \citet{BelskayaShevchenko00}. For the dust 
   particles ejected, we assumed the same values of $p_v$ and $b$ as 
   for the asteroid. 

     The remaining model parameters are the event times and duration,
     the dust masses ejected, and the terminal particle
     speeds. The speed is assumed to follow a law of the kind
     $v=v_0\beta^{\gamma}$. For a sublimating body in which
       particles are accelerated by gas drag, $\gamma\sim$0.5. For a
       simple model of 
       rotational disruption, a size-independent speed would be
       expected, so that $\gamma\sim$0. The event times are
       preliminarily guessed by a syndyne-synchrone 
     analysis (Finson \& Probstein \citeyear{Finson}), which applies to zero
     ejection velocity conditions, and then refined by the Monte
     Carlo dust tail model. The dust loss rates are assumed constant
     during the occurrence of the ejection events.  

     The fitting procedure is accomplished by performing a
     minimisation of the quantity 
     $\sigma=\sum\limits_{l=1}^{l=N} \sigma_l$, where the summation is
     extended to all the images included in the analysis, and 
     $\sigma_l=\sqrt{\sum\limits_{i,j}
         \frac{(log_{10}[B_{obs}(i,j)]-log_{10}[B_{mod}(i,j)])^2}{N_l}}$ is the 
     standard deviation of the model brightness ($B_{mod}$) with respect to
     the observed brightness($B_{obs}$) for image $l$, where $(i,j)$
     represent the image pixel location, and $N_l$ the total number of
     pixels of image $l$. This minimization is performed using a
     multidimensional downhill simplex method
     (Nelder \& Mead \citeyear{NelderMead65}). 

   \begin{figure*}
   \centering
   \includegraphics[width=15cm]{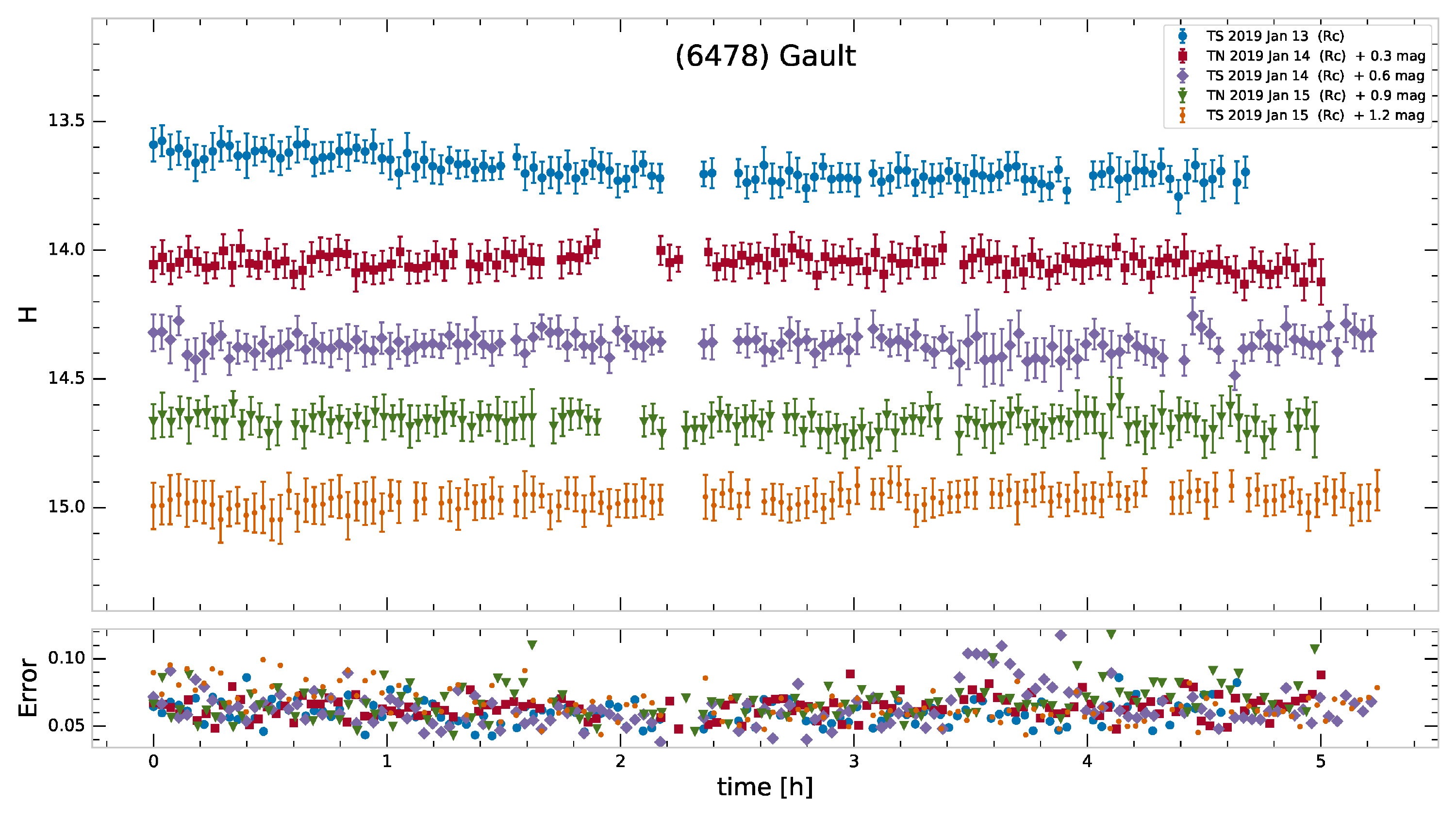}
      \caption{Absolute $H$ magnitude lightcurves of (6478) Gault obtained
        with TRAPPIST-North (TN) and -South (TS) on 2019 January
        13, 14 and 15. The TS 2019 January 13 data correspond to the
        true magnitude values 
        on the y-axis, while the other datasets are shifted by
        multiples of 0.3 mag downwards, as indicated. Errors
        correspond to 2$\sigma$.}
         \label{LC}
   \end{figure*}

\section{The Lightcurve}

The aim of obtaining the lightcurves was primarily to search for the rotation
period of Gault. The apparent $R_c$ magnitudes were converted to
absolute magnitudes $H$ by assuming $V-R_c$=0.49 (Shevchenko \& Lupishko
\citeyear{Shevchenko98}), and a slope parameter $G$=0.227
(Warner et al. \citeyear{Warner09}), both values appropriate for S-type
asteroids. The resulting lightcurves are shown in Figure 2. Each
lightcurve is essentially flat, except for a weak slope in the first
half of the lightcurve of January 13. All these data did not allow us to
determine a rotation period or to observe any trend of the magnitude on
several consecutive days. If the rotation period of Gault was an
integer divider of 24 h, we would have observed the same part of the
rotation each time. However, as we observed 3 nights continuously for
5 h, 7.3 h and 7.3 h, respectively, it is very unlikely that we were
observing each time the same flat part of its lightcurve.  The
possibilities to explain a flat lightcurve for an asteroid include a
nearly spherical shape, pole-on observation, and the presence of dust
around the asteroid hiding its surface.  Fast rotators typically
present a roughly spheroidal shape, resembling a spinning top. The
amplitude of the lightcurves of this kind of asteroids is small
(i.e., $\lesssim$0.2 mag). If
the object is embedded in a coma this small variation can easily be
masked (Licandro et al. \citeyear{Licandro00}). For Gault,
given the magnitude of
the inactive asteroid ($H$=14.4), we find that approximately half of
the total brightness is produced by the dust envelope on the 
lightcurve dates. A recent study performed by \citet{Kleyna19}
points to a $\sim$2h rotation period for Gault, which would indicate a
superfast rotator and probable rotational disruption as the cause of
the dust ejection. Further photometric 
observations when the asteroid will have no detectable coma can give
additional results on the rotation properties.  
     
\section{Results and Discussion}

Since the number of free parameters in the model is large, in order to
make the problem tractable we had to
perform preliminary searches of some of the input parameters. Thus,
the exponent $\gamma$ in the particle ejection speed equation was set 
to $\gamma$=0.5. Although this value is 
       appropriate to gas drag acceleration from an ice-sublimating
       body, we had to assume it in order to 
       explain the observed widening of
     the main tail as a function of 
     distance to the asteroid, mainly for the latest images acquired 
     (see Figure 1b and Figures A.8 to A.10). At
     present, we do not find any explanation for this fact, as 
     ice sublimation seems rather improbable. The overall 
     width of the tails depend on 
     $v_0$. We found a value $v_0$= 3 m s$^{-1}$ as appropriate, which
     implies terminal speeds values of $\sim$0.4 m s$^{-1}$ for
     10 $\mu$m particles, and lower for larger particles. These
     terminal speeds are comparable to the escape 
     velocity of the asteroid at distances where the asteroid gravity
     becomes negligible compared to solar gravity. Thus, if we assume
     for the bulk density of the asteroid a 
     value typical for S-type ($\rho_{bulk}$=2710 kg m$^{-3}$,
     Krasinsky et al. 
     \citeyear{Krasinsky02}), the escape velocity at 100 km from the
     asteroid surface would be $\sim$0.4 m s$^{-1}$, where the ratio
     of solar gravity to asteroid gravity is larger than 10$^3$. This
     indicates that the particles populating Gault tails were likely
     released at near escape speeds from the asteroid surface. This would be
     compatible with a rotational breakup. However, the speed
       dependence on particle size seems incompatible with this
       mechanism. A simple model of rotational disruption would yield a 
       size-independent ejection speed.
     
     On the other hand, the event times are preliminarily guessed by a
     syndyne-synchrone 
     analysis (Finson \& Probstein \citeyear{Finson}), which applies to zero
     ejection velocity conditions, and then refined by the Monte
     Carlo dust tail model. The duration of such
     events is only weakly constrained by the model,  
     provided they are shorter than $\sim$5 days. We then fixed the duration
     to 1 day for simplicity, but keeping in mind that this can be in
     fact much shorter.

  After these preliminary 
     searches of the mentioned parameters, we then solved for the
     total mass of dust 
     ejected, and the power-law index of the differential size
     distribution. It soon 
     became apparent that a single power-law function was inadequate to
     describe the variation of the brightness of the main tail with
     distance to the asteroid. It comes from the fact that
     the brightness 
     along the tail first slightly increases or keeps constant tailward, and
     finally decreases 
     (see Figure 1,  and Figures A.1-A.10). In order to fit properly the
     observed brightness profile we adopted a broken power-law with
     two bending points. This left us with a total of seven 
     fitting parameters: the two bending points, the three power-law
     indexes, and the dust mass released during each event. 
 The resulting best-fitted differential size distribution was characterised by 
 power-law indexes of --2.28 (between 1
 $\mu$m and 15 $\mu$m),  --3.95 (between 15 $\mu$m and 870 $\mu$m), and
 --4.22 (between 870 $\mu$m and 1 cm).  This probably reflects the
   asteroid regolith size distribution. Broken power-law functions are
 common when describing the size distributions of asteroid dust bands
 (Nesvorn\'y et al. \citeyear{Nesvorny06}), the boulder distribution 
 in asteroid regoliths (Tancredi et 
 al. \citeyear{Tancredi15}), or in cometary dust (Fulle et
 al. \citeyear{Fulle16}, Moreno et al. \citeyear{Moreno16}), always with
 a tendency to increase slope with size.

 The total dust mass released
 were of 1.4$\times$10$^7$, and 1.6$\times$10$^6$ kg, for the 
 2018 November 5, and 2019 January 2 events, respectively. We
 need to emphasise that these masses are lower limits to the actual
 total dust mass 
 ejected. Thus, should a few large and massive boulders have
   been ejected, they
 would not contribute significantly to the brightness, but would do to
 the mass. We underline that the same differential  
 size distribution is assumed for the two ejection events.  

 It is very interesting to note the remarkably similar size
 distribution, ejected masses, and velocities recently reported by
 \citet{Ye19}. 
 
The fits to the data are shown in Figures A.1-A.10 in Appendix
1. These fits are provided as a synthetic image which best fits
the observed image at a given date, an isophote map showing the
measured and the modelled isophotes near the asteroid location, 
and the brightness along the tail,
for both the observation and the model.  The ``bumps'' along some of
these measured profiles (most evident in Figures A.4, A.5, A.7, A.9, and A.10)
are due to contamination by field stars.   
The asteroid contribution to the
brightness profile near the optocenter is more and more important as
time progresses, as the dust is being blown away 
by radiation pressure.

 The total mass ejected, 1.56$\times$10$^{7}$ kg,  is negligible
  compared to the asteroid mass (1.3$\times$10$^{14}$ kg), and would
  correspond to a spherical volume of just $\sim$10 m radius.

\section{Conclusions}
We have carried out observations and models to characterise the dust
properties and event timelines of the double-tailed active
asteroid Gault. The following conclusions can be drawn:  

   \begin{enumerate}
     
      \item Asteroid Gault has experienced two short-lived activity periods
        of a maximum duration of 5 days, but might be much shorter,
        separated by a period of inactivity of nearly two months. A
        similar activity pattern has been previously observed for  
        inner belt asteroid 311P.   

      \item The total dust masses ejected for particles in the 1
        $\mu$m to 1 cm radius were 1.4$\times$10$^7$, and
        1.6$\times$10$^6$ kg, for the November 5, 2018, and January 2,
        2019 events, respectively. These masses represent a negligible 
          fraction of the asteroid mass, estimated at
          1.3$\times$10$^{14}$ kg. 

      \item To fit the brightness distribution along the main tail, the
        differential size distribution function can be described as a broken 
        power-law function, with power indexes of --2.28, --3.95, and
        -4.22, and bending points near 15 $\mu$m and 870 $\mu$m. 

      \item  The ejection speeds are found to be close to the escape speed
        of Gault, a fact compatible with rotational
        disruption phenomena. However, the model results point to a  
        $\beta^{0.5}$ dependence of the speeds, typical of 
        sublimation-driven processes. This mechanism is, however, very
        unlikely owing to the inner belt character of the
        object. Only a detailed lightcurve of the naked, dust-free,
        asteroid in combination with dynamical 
        modelling might shed additional light on the ejection mechanism. 

      \end{enumerate}

   \begin{acknowledgements}

 We would like to thank the reviewer for his/her constructive comments which
helped to substantially improve the quality of the paper.
     
  F. Moreno acknowledges financial support from the State Agency for
  Research of the Spanish MCIU through the "Center of Excellence
  Severo Ochoa" award for the Instituto de Astrof\'\i sica de
  Andaluc\'\i a (SEV-2017-0709). 

 This article is partially based on observations made with the Gran Telescopio
Canarias, installed in the Spanish Observatorio del Roque de los 
Muchachos of the Instituto de Astrof\'\i sica de Canarias (IAC), in the 
island of La Palma and on observations made in the Observatorios de
Canarias del IAC with the MUSCAT2 instrument on the Carlos Sanchez
telescope operated on the island of Tenerife by the IAC in the
Observatorio del Teide.

TRAPPIST-North is a project funded by the
University of Li\`ege, in collaboration with Cadi Ayyad University of
Marrakech (Morocco). TRAPPIST-South is a project funded by the Belgian
Fonds (National) de la Recherche Scientifique (F.R.S.-FNRS) under
grant FRFC 2.5.594.09.F. E.J and M.G are F.R.S.-FNRS Senior Research
Associates.  

This work is also based on observations obtained at the Southern
Astrophysical Research (SOAR) telescope, which is a joint project of
the Minist\'erio da Ci\^encia, Tecnologia, e Inovaç\~ao
(MCTI) da Rep\'ublica Federativa do Brasil, the U.S. National Optical
Astronomy Observatory (NOAO),
the University of North Carolina at Chapel Hill (UNC), and Michigan State University (MSU).

This work was supported by contracts AYA2015-67152-R and AYA
2015-71975-REDT from the Spanish Ministerio de Econom\'\i a y
Competitividad (MINECO). J. Licandro gratefully acknowledges support
from contract AYA2015-67772-R (MINECO, Spain).
\end{acknowledgements}

%

\begin{thebibliography}{}
  \bibitem[Belskaya \& Shevchenko(2000)]{BelskayaShevchenko00} Belskaya, I.N., \& Shevchenko,
    V.G. 2000, Icarus, 147, 94

  \bibitem[Burns(1979)]{Burns79} Burns, J.A., Lamy, P.L., Soter, S. 1979,
    Icarus, 40, 1


  \bibitem[Chambers(1999)]{Chambers99} Chambers, J.E., 1999, \mnras,
    304, 793

    
  \bibitem[Finson \& Probstein(1968)]{Finson} Finson, M.L., \&
    Probstein, R.F. 1968, \apj, 
    154, 353

  \bibitem[Fulle(1989)]{Fulle89} Fulle, M. 1989, A\&A, 217, 283

  \bibitem[Fulle et al.(2016)]{Fulle16} Fulle, M., Marzari, F., Della
    Corte, V., et al. 2016, ApJ, 821, 19

  \bibitem[Hainaut et al.(2014)]{Hainaut14} Hainaut, O.R., Boehnhardt,
    H., Snodgrass, 
    C., et al. 2014, A\&A, 563, A75



  \bibitem[Harris \& Lagerros(2002)]{HarrisLagerros02} Harris, A.W.,
    \& Lagerros, 
    J.S.V. in Asteroids III, Univ. of Arizona Press, Tucson

  \bibitem[Haghighipour(2009)]{Haghighipour09} Haghighipour, N., 2009,
    Meteor. and 
    Planet. Sci., 44,1863


  \bibitem[Jehin et al.(2011)]{Jehin11} Jehin, E., Gillon, M., Queloz,
    D., et al. 2011, The Messenger, 145
    
  \bibitem[Jehin(2019)]{Jehin19} Jehin, E., Ferrais, M., Moulane, Y.,
    et al. 2019, Central Bureau Electronic Telegrams, 4606 



  \bibitem[Jewitt et al.(2009)]{Jewitt09} Jewitt, D., Yang, B.,
    Haghighipour, N. 2009, 
    \aj, 137, 4313


  \bibitem[Jewitt et al.(2013)]{Jewitt13} Jewitt, D., Agarwal, J.,
    Weaver, H., et 
    al. 2013, \apj, 778, L21


  \bibitem[Jewitt et al.(2015)]{Jewitt15} Jewitt, D., Hsieh, H., and Agarwal,
    J. 2015, in Asteroids IV, Univ. of Arizona Press, Tucson

  \bibitem[Jewitt et al.(2018)]{Jewitt18} Jewitt, D., Weaver, H.,
    Mutchler, M., 2018, 
    \aj, 155, 231

    
\bibitem[Kleyna et al.(2019)]{Kleyna19} Kleyna, J.T., Hainaut, O.R.,
  Meech, K.J., et al. 2019, ApJ Lett., in press
    
\bibitem[Krasinsky et al.(2002)]{Krasinsky02} Krasinsky, G.A.,
  Pitjeva, E.V., Vasilyev, M.V. et al. 2002, Icarus, 158, 98 

\bibitem[Licandro et al.(2000)]{Licandro00} Licandro, J.,
  Serra-Ricart, M., Oscoz, A. et
    al. 2000, Astron. J., 119, 3133

  \bibitem[Luu \& Jewitt(1989)]{LuuJewitt89} Luu, J., \& Jewitt, D.,
    1989, Astron. J., 98, 1905

  
  \bibitem[Mommert(2017)]{Mommert17} Mommert, M. 2017, Astronomy and Computing,
    18, 47

  \bibitem[Moreno et al.(2014)]{Moreno14} Moreno, F., Licandro, J.,
    \'Alvarez-Iglesias, C., et al. 2014, \apj, 781, 118

    
  \bibitem[Moreno et al.(2016)]{Moreno16} Moreno, F., Snodgrass, C.,
    Hainaut, O., et 
    al., 2016, A\&A, 587, 155

  \bibitem[Moreno et al.(2017)]{Moreno17} Moreno, F., Pozuelos, F.J.,
    Novakovi\'c, B., et al., 2017, \apj, 837, L3

  \bibitem[Nelder \& Mead(1965)]{NelderMead65} Nelder, J.A., \& Mead,
    R. 1965, Computer Journal, 7, 308
	
\bibitem[Nesvorn\'y et al.(2006)]{Nesvorny06} Nesvorn\'y, D.,
  Vokrouhlick\'y, D., Bottke, W.F. et al. 2006, Icarus, 181, 107

  \bibitem[Smith et al.(2019)]{Smith19} Smith, K.W., Denneau, L.,
    Vincent, J.B., et 
    al. 2019, CBET 4594.

  \bibitem[Shevchenko \& Lupishko(1998)]{Shevchenko98} Shevchenko,
    V.G., \& Lupishko, D.F. 1998, Solar Syst. Res. 32, 220

  \bibitem[Tancredi et al.(2015)]{Tancredi15} Tancredi, G., Roland,
    S., and Bruzzone, S. 2015, Icarus, 247, 279

\bibitem[Tsuchiyama et al.(2011)]{Tsuchiyama11} Tsuchiyama, A.,
  Uesugi, M., Matsushima, 
  T., et al. 2011, Science, 333, 1125
   
 \bibitem[Warner et al.(2009)]{Warner09} Warner, B.D.,  Harris, A.W.,
   Pravec, P. 2009, Icarus, 202, 134 
 
 \bibitem[Ye et al.(2019)]{Ye19} Ye, Q., Kelley, M., Bodewits, D., et
   al. 2019, ApJ Lett., in press.



   

\end{thebibliography}
%

\begin{appendix}
\section{Monte Carlo dust tail model fits to the images}
  In this Appendix we show the model fits for all the images shown in
  Table 1.
  \begin{figure*}
   \centering
   \includegraphics[angle=0,width=15cm]{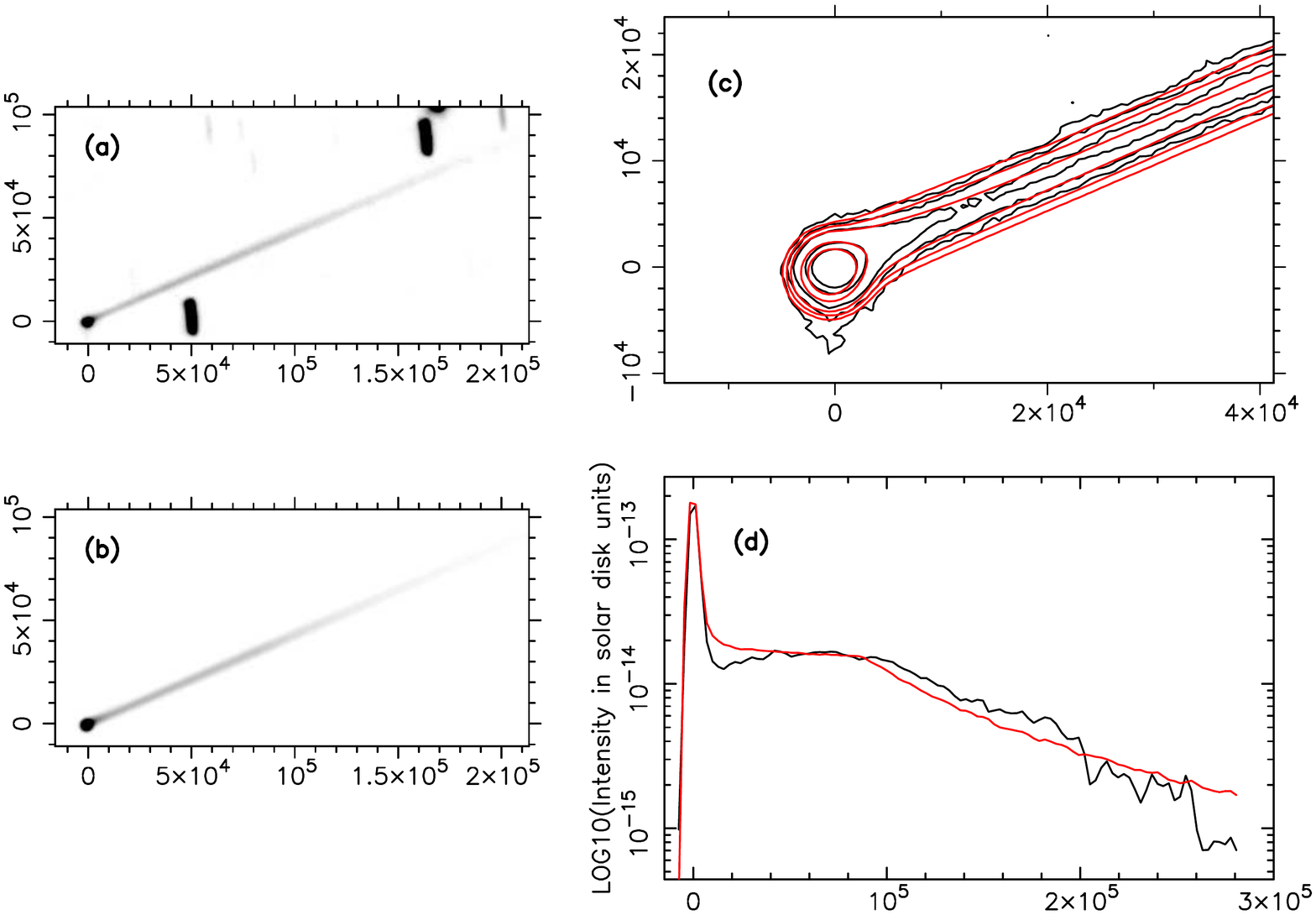}
      \caption{Results of the model fit for the image obtained with
        TCS on 2019 January 11. Panel (a) is the observed image. Panel
        (b) is the 
        model image, using the same brightness scale as in (a).  Panel (c) shows
        the isophote field near the 
        asteroid location (observation in black contours, and model in
        red contours). In Panel (d), a comparison between the
        observed intensity along the main tail (black line), and
        the model (red line), is given. Axes are labelled in km
        projected on the sky at the asteroid distance. In
        panels (a)-(c), North is up, East 
        to the left.}
         \label{Graph1}
   \end{figure*}

  \begin{figure*}
   \centering
   \includegraphics[angle=0,width=15cm]{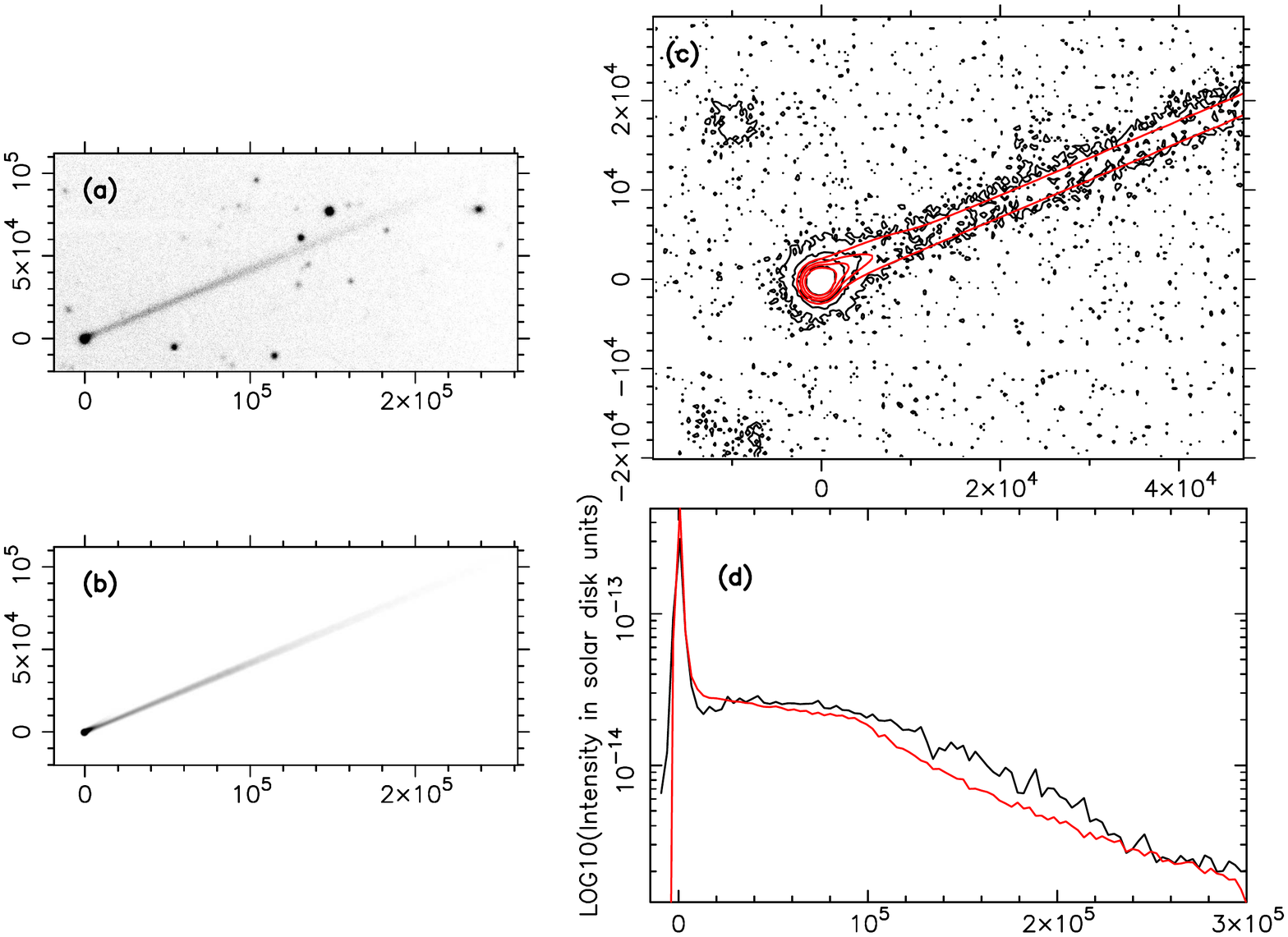}
      \caption{As Figure A.1, but for the 
        GTC image on 2019 January 13. }
         \label{Graph2}
  \end{figure*}

  \begin{figure*}
   \centering
   \includegraphics[angle=0,width=15cm]{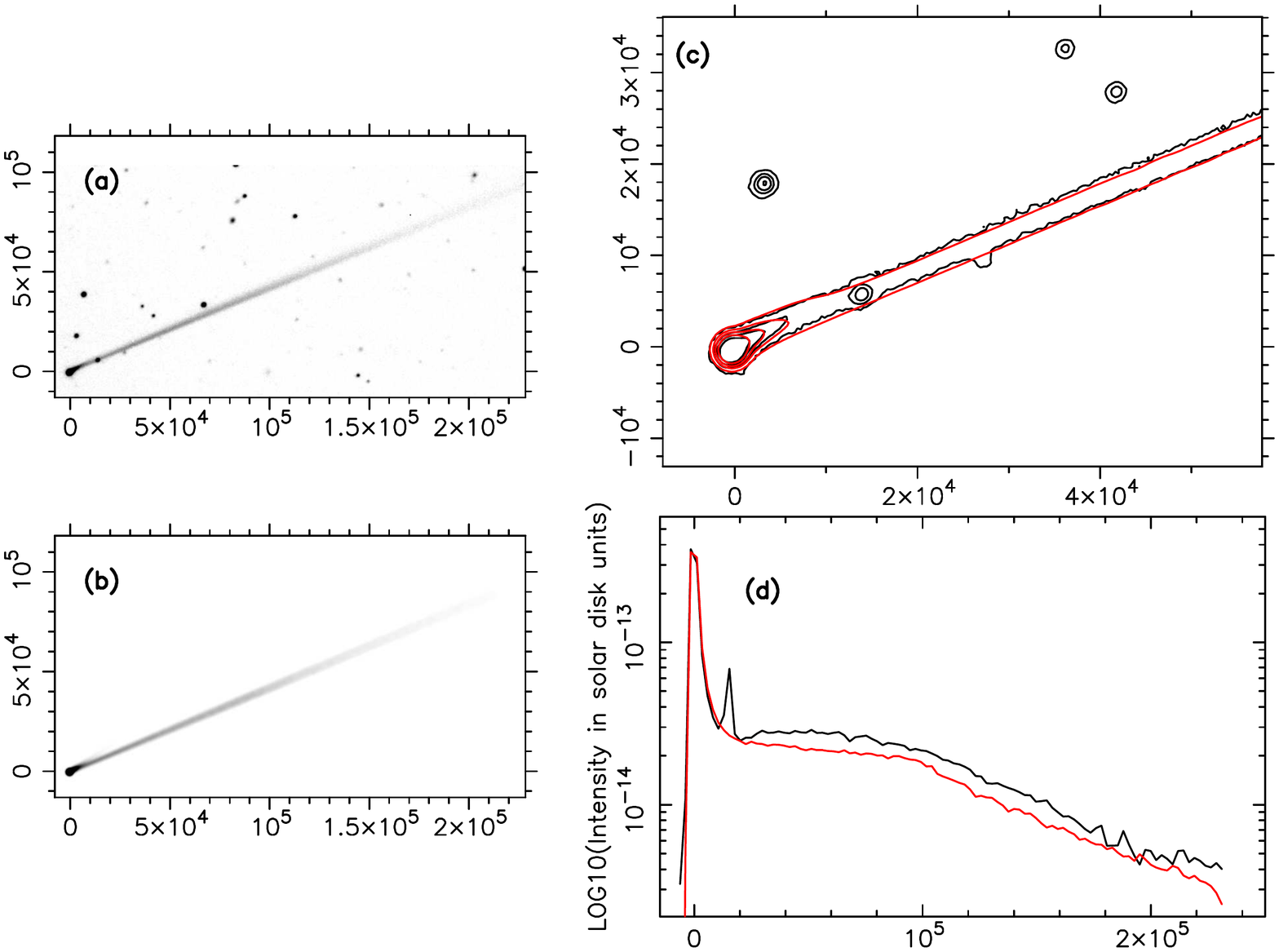}
      \caption{As Figure A.1, but for the 
        GTC image on 2019 January 14.}
         \label{Graph3}
  \end{figure*}

  \begin{figure*}
   \centering
   \includegraphics[angle=0,width=15cm]{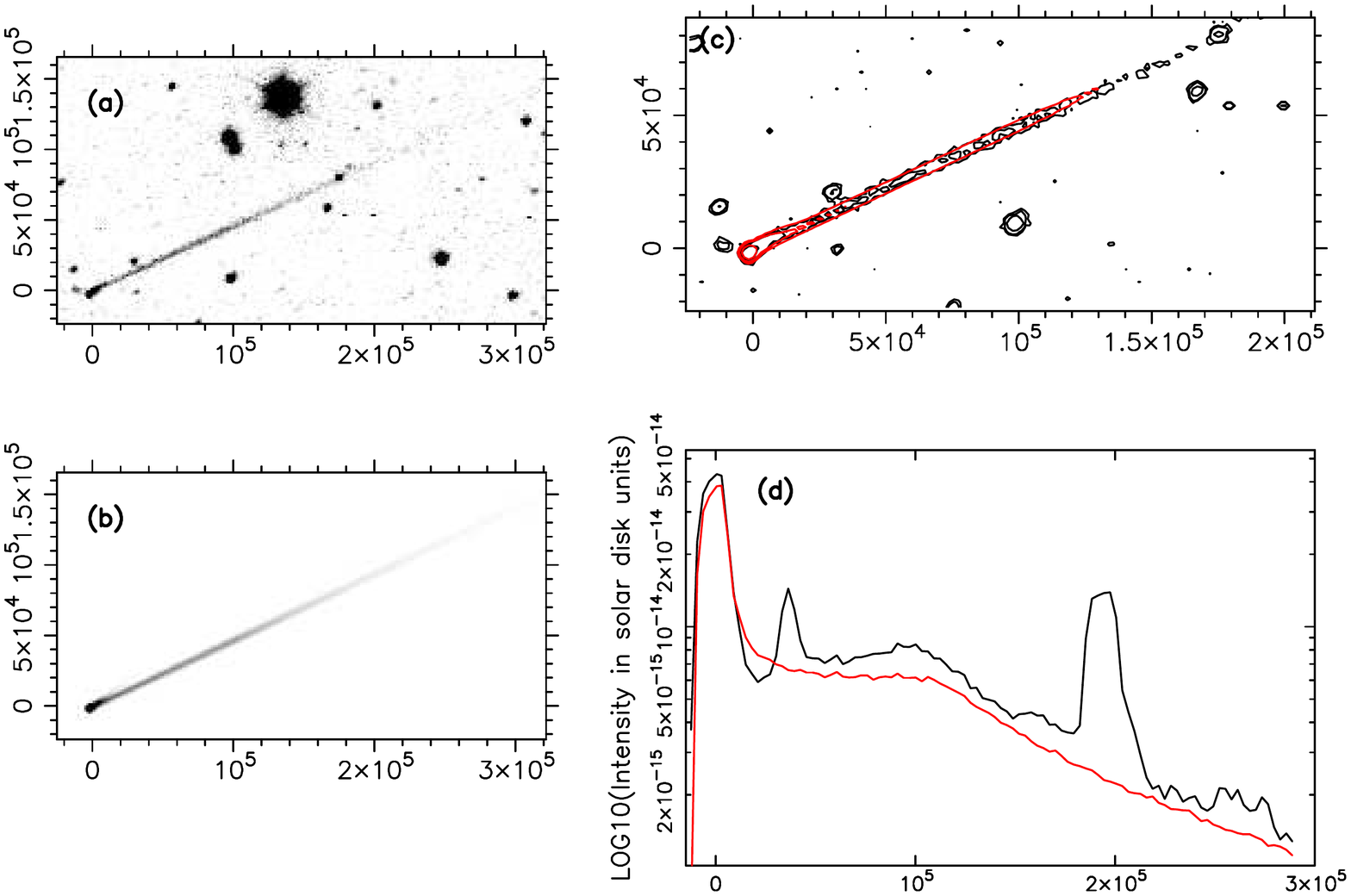}
      \caption{As Figure A.1, but for the 
        TRAPPIST image on 2019 January 21.}
         \label{Graph4}
  \end{figure*}

  \begin{figure*}
   \centering
   \includegraphics[angle=0,width=15cm]{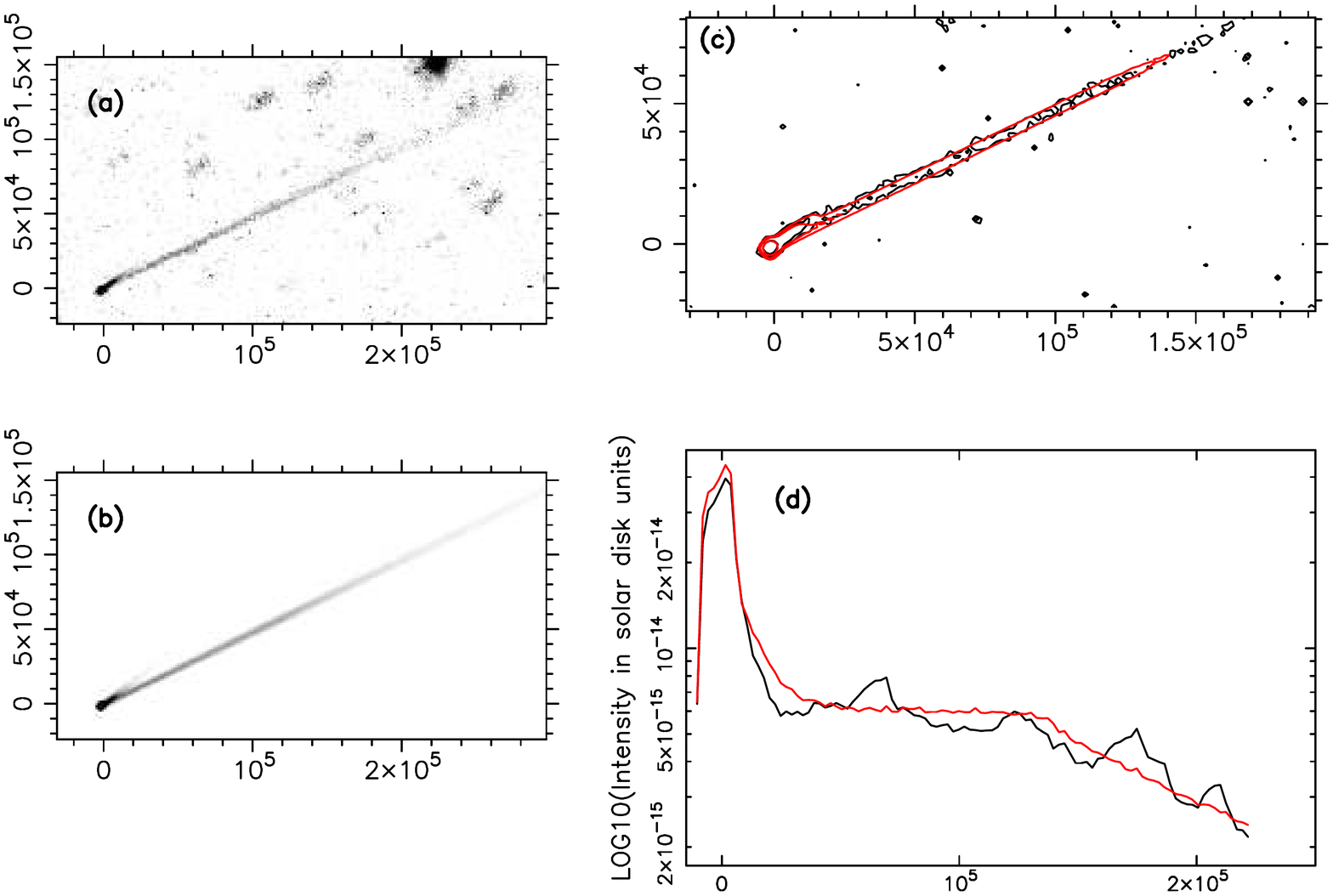}
      \caption{As Figure A.1, but for the 
        TRAPPIST image on 2019 January 29.}
         \label{Graph5}
  \end{figure*}

   \begin{figure*}
   \centering
   \includegraphics[angle=0,width=15cm]{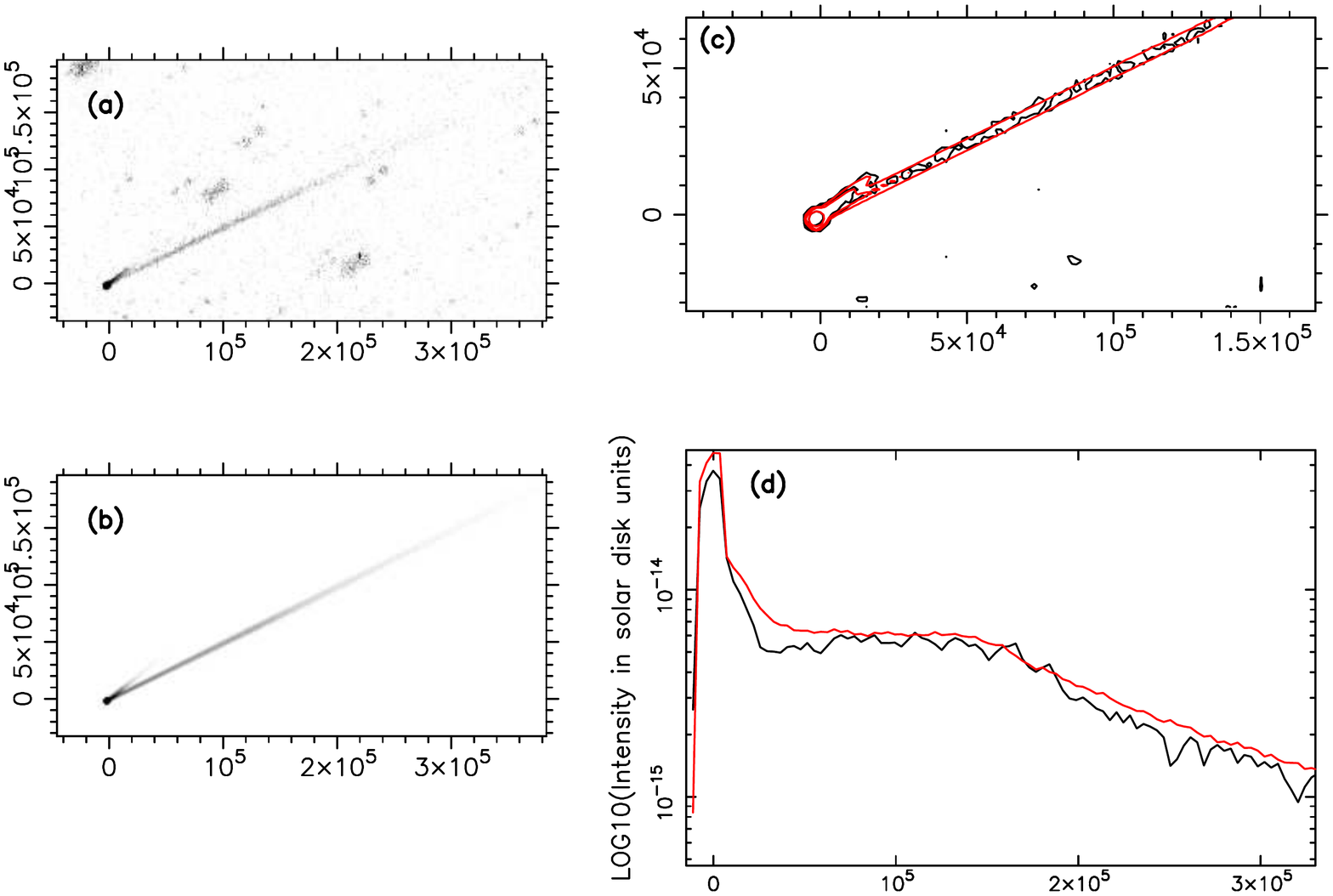}
      \caption{As Figure A.1, but for the 
        TRAPPIST image on 2019 February 5.}
         \label{Graph6}
  \end{figure*}

  \begin{figure*}
   \centering
   \includegraphics[angle=0,width=15cm]{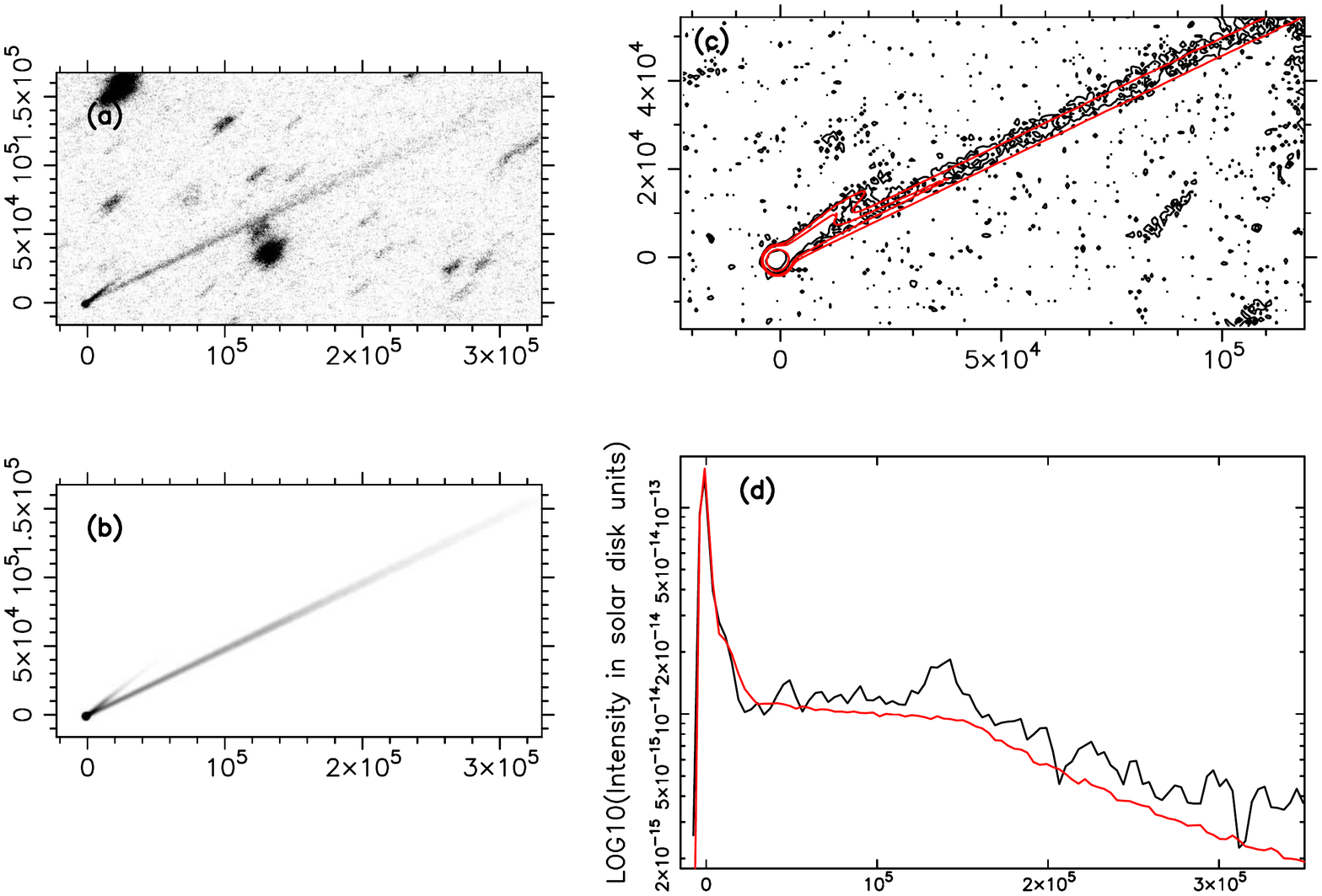}
      \caption{As Figure A.1, but for the 
        TRAPPIST image on 2019 February 7.}
         \label{Graph7}
  \end{figure*}

  \begin{figure*}
   \centering
   \includegraphics[angle=0,width=15cm]{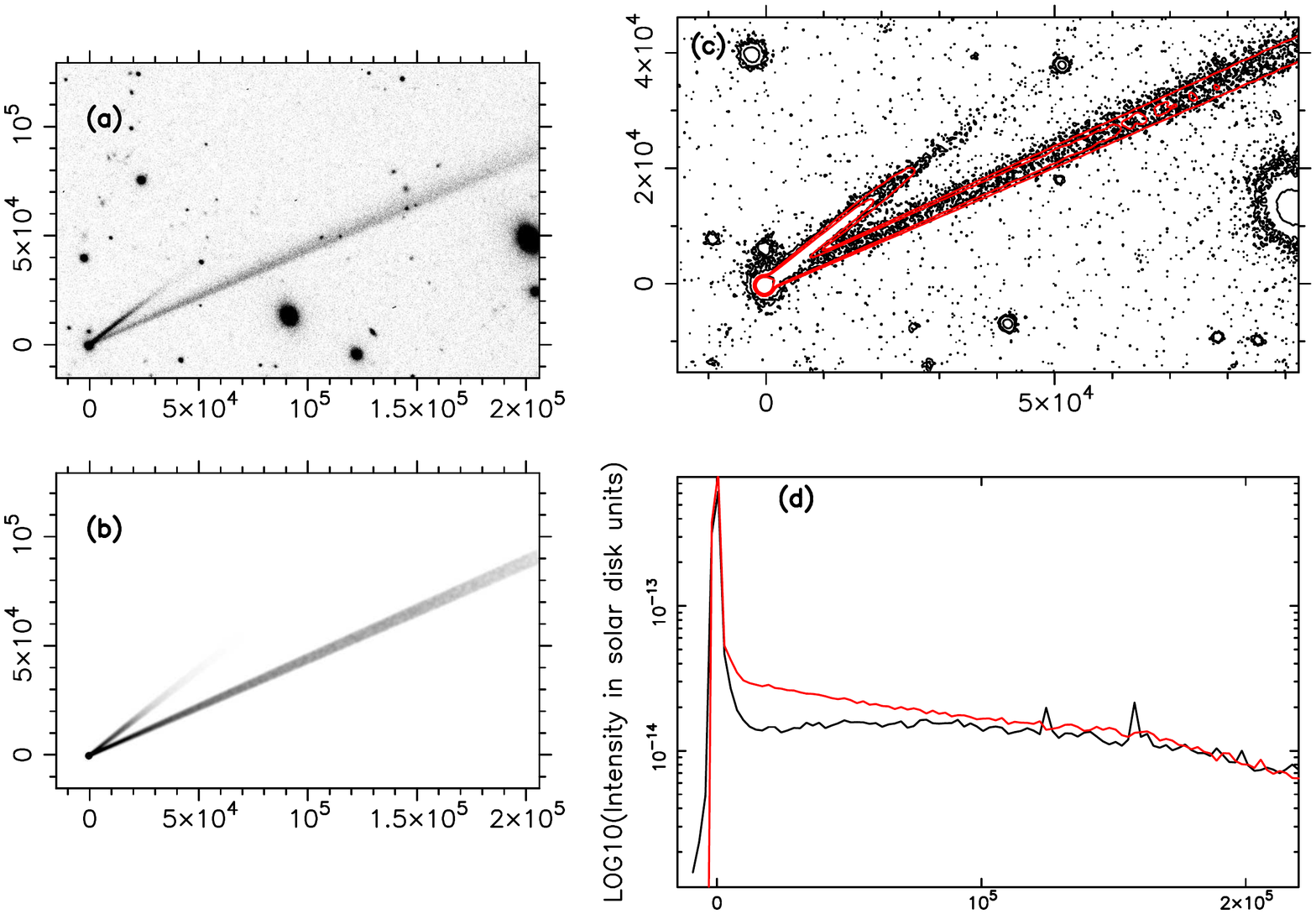}
      \caption{As Figure A.1, but for the 
        SOAR image on 2019 February 15.}
         \label{Graph8}
  \end{figure*}

  \begin{figure*}
   \centering
   \includegraphics[angle=0,width=15cm]{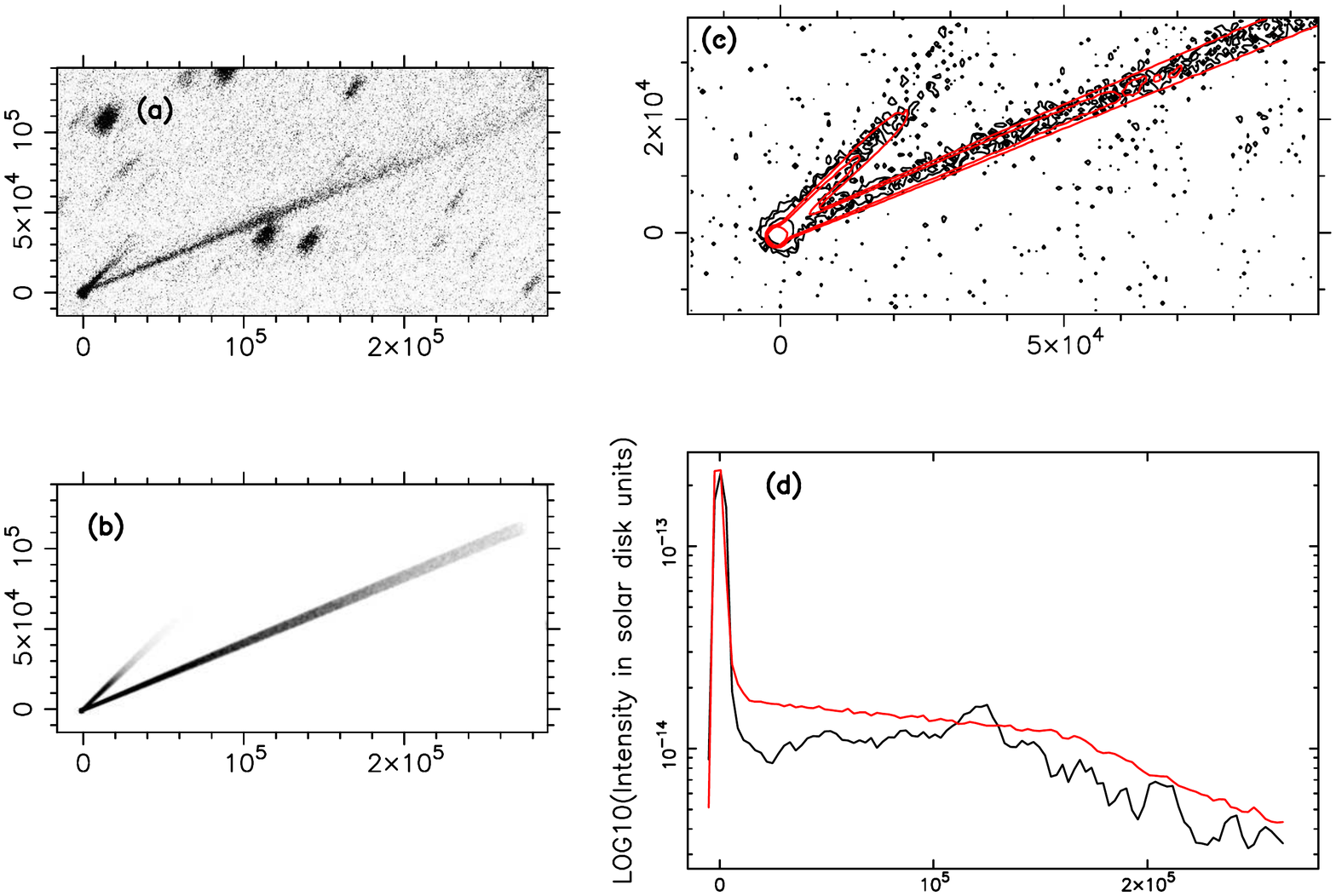}
      \caption{As Figure A.1, but for the 
        TRAPPIST image on 2019 March 5.}
         \label{Graph9}
  \end{figure*}

    \begin{figure*}
   \centering
   \includegraphics[angle=0,width=15cm]{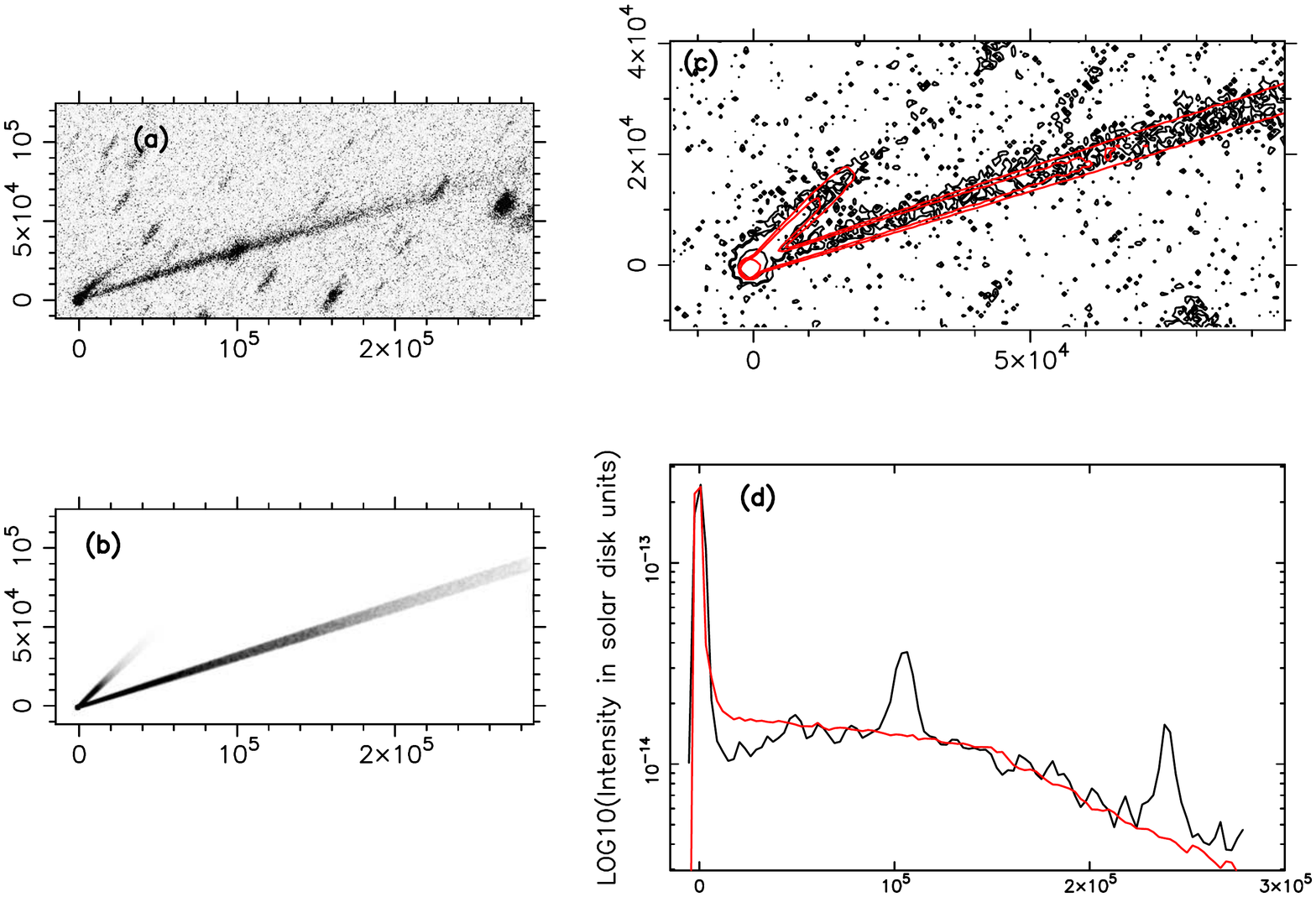}
      \caption{As Figure A.1, but for the 
        TRAPPIST image on 2019 March 13.}
         \label{Graph11}
  \end{figure*}

\end{appendix}

\end{document}